\documentclass[10pt,conference]{IEEEtran}
\usepackage{cite}
\usepackage{amsmath,amssymb,amsfonts}
\usepackage{algorithmic}
\usepackage{graphicx}
\usepackage{textcomp}
\usepackage{xcolor}
\usepackage[hyphens]{url}
\usepackage{fancyhdr}
\usepackage{hyperref}
\usepackage{tabularx}
\usepackage{booktabs}
\usepackage{array}
\usepackage{listings}
\usepackage[dvipsnames]{xcolor}
\usepackage{tikz}
\usepackage{enumitem}
\usepackage{comment}
\usepackage{float}
\usepackage{subcaption}

\definecolor{MyDarkGreen}{rgb}{0,0.8,0.0}
\definecolor{MyDarkBlue}{rgb}{0,0,0.8}
\definecolor{MyDarkRed}{rgb}{0.6,0,0.0}

\title{Divide and conquer: Scalable performance and energy in MCM GPUs}

\def\hpcacameraready{} 

\newcommand\hpcaauthors{Mario Ibáñez Bolado$\dagger$, Borja Pérez Pavón$\dagger$, Jose Luis Bosque Orero$\dagger$, Julio Ramon Beivide$\dagger$}
\newcommand\hpcaaffiliation{Universidad de Cantabria$\dagger$}
\newcommand\hpcaemail{\{ibanezbm, perezpavonb, bosquejl, beividej\}@unican.es}

\author{
  \ifdefined\hpcacameraready
    \IEEEauthorblockN{\hpcaauthors{}}
      \IEEEauthorblockA{
        \hpcaaffiliation{} \\
        \hpcaemail{}
      }
  \fi 
}

\fancypagestyle{camerareadyfirstpage}{%
  \fancyhead{}
  
  \fancyhead[C]{
    \ifdefined\aeopen
    \parbox[][12mm][t]{13.5cm}{}    
    \else
      \ifdefined\aereviewed
      \parbox[][12mm][t]{13.5cm}{}
      \else
      \ifdefined\aereproduced
      \parbox[][12mm][t]{13.5cm}{}
      \else
      \parbox[][0mm][t]{13.5cm}{}
    \fi 
    \fi 
    \fi 
    \ifdefined\aeopen 
      \includegraphics[width=12mm,height=12mm]{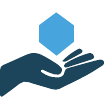}
    \fi 
    \ifdefined\aereviewed
      \includegraphics[width=12mm,height=12mm]{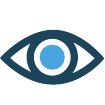}
    \fi 
    \ifdefined\aereproduced
      \includegraphics[width=12mm,height=12mm]{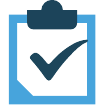}
    \fi
  }
  \fancyfoot[C]{}
}
\begin{document}
\maketitle

\ifdefined\hpcacameraready 
  \thispagestyle{camerareadyfirstpage}
  \pagestyle{empty}
\else
  \thispagestyle{plain}
  \pagestyle{plain}
\fi

\newcommand{\hpcaheight}{0mm}
\ifdefined\eaopen
\renewcommand{\hpcaheight}{12mm}
\fi


\begin{abstract}

Multi-chip-module (MCM) GPUs offer a promising path to scale compute capability beyond monolithic designs by integrating multiple chiplets on a common package. However, the impact of disaggregation on performance scalability and energy consumption remains underexplored. The design space grows rapidly across dimensions such as SMs per chiplet, chiplet count, and interconnection network. The inter-chiplet network is particularly critical, as it determines whether additional compute resources translate into performance gains. This limited understanding leaves industry and research without clear guidance on the performance and energy trade-offs of MCM GPU scaling. In this work, we investigate whether distributing compute and memory capability across multiple chiplets offers a more scalable alternative to concentrating resources. We quantify their effects on performance, energy, and efficiency and examine how inter-chiplet topology influences scalability at different system sizes. Our results demonstrate that a 16 chiplet Torus configuration with 256 SMs delivers a remarkable $2.40\times$ performance improvement over a state-of-the-art MCM architecture with the same compute capability, while simultaneously reducing energy consumption by $4.45\times$. These substantial gains provide evidence that disaggregation is a first-order architectural factor and will be critical to unlocking the performance and energy-efficiency potential of next-generation GPUs.
\end{abstract}

\section{Introduction}

High-Performance Computing (HPC) and Artificial Intelligence (AI) applications increasingly rely on GPUs as essential computational accelerators \cite{vijayaraghavan2017design, villa2014scaling}. Their adoption is largely driven by their ability to execute a large number of operations concurrently through highly parallel execution models, as well as their excellent energy efficiency. Increasing the resources available to a GPU system can substantially improve its ability to support larger and more demanding workloads. Previous research has shown that many applications scale effectively as additional processing units and computational capability are incorporated \cite{jouppi2017datacenter, lee2010debunking, Perez:2021}. HPC and AI systems have become capable of tackling larger and more challenging problems, ranging from complex scientific simulations to sophisticated learning models and data-analysis workloads.

In this context, Multi-Chip-Module GPUs \cite{arunkumar2017mcm, arunkumar2019understanding}, or MCM GPUs were conceived to overcome the physical and economic limitations that make the continued scaling of monolithic GPU designs increasingly impractical. This shift is driven by two converging limitations: the slowdown of traditional transistor scaling as Moore’s Law approaches its practical limits, and the “area wall”, which makes further enlargement of a monolithic die increasingly difficult and costly \cite{waldrop2016chips, esmaeilzadeh2011dark, frank2001device, hennessy2017computer, areaWall}.

To address these limitations, chiplet architectures \cite{ma2022survey, smith202411, lee2016overview} have emerged as a promising alternative to traditional monolithic designs for both CPUs and GPUs. In the GPU domain, this approach gives rise to MCM GPUs, which combine multiple GPU chiplets within a single package \cite{arunkumar2017mcm}. Each GPU chiplet contains a subset of the compute, memory, and communication resources of the complete GPU. The GPU chiplets communicate using an inter-chiplet network, that enables data exchange and coordination among the modules within the package. This modular organization helps overcome the scalability, manufacturing, and cost limitations of monolithic designs. MCM architectures can provide higher compute density and better scalability than single die designs \cite{arunkumar2019understanding}.

Despite their scalability potential, MCM GPUs introduce an architectural trade off regarding how additional resources are incorporated within the package. A \textbf{scale-out} policy increases the computational capabilities of an MCM GPU architecture by adding more chiplets while each one contains a predetermined number of SMs and associated resources, all of them connected through a chiplet network. This approach provides greater aggregate compute capability by incorporating additional GPU chiplets, but its scalability is limited by package area, interconnect bandwidth, communication overhead, and power delivery. Alternatively, a \textbf{scale-up} strategy increases the capabilities of each GPU chiplet by incorporating additional resources, such as more compute units or memory resources. However, this approach is  constrained by chiplet area, manufacturing yield, power consumption, and thermal limits. Finally, \textbf{hybrid} strategies combine both approaches by increasing the resources within each chiplet while also increasing the number of chiplets integrated within the package.



Scalability must also be evaluated from an energy perspective. Although increasing the number of chiplets may improve throughput, communication and data movement consume additional energy and may reduce overall resource utilization. The resulting performance gains must therefore be weighed against the power consumed by the interconnect, and packaging infrastructure. Concluding whether scale-out or scale-up provides the best performance and energy efficiency is thus a central design challenge for MCM GPU systems.

Consequently, the effectiveness of MCM GPUs cannot be assessed solely by their aggregate compute capability. A comprehensive evaluation of MCM GPU scalability must therefore quantify how these factors affect execution time and energy consumption as the system grows. In this work, we pose the divide and conquer premise: disaggregation enables more effective scaling of MCM GPUs. We evaluate this principle across different scaling strategies and analyze its impact on performance, energy, and efficiency, seeking to guide future research and industry product development. Our results show that disaggregating the system into 16 chiplets torus can yield performance improvements of up to $2.40\times$,  while reducing energy consumption by $4.45\times$, assuming adequate communication facilities are provided, even at scales of only 256 SMs, which may be reasonably expected in upcoming GPU generations. Moreover, highly disaggregated designs can match or outperform more compact configurations equipped with $2\times$ or $4\times$ as many compute resources, demonstrating that resource distribution can be more important than compute capability alone.


The main contributions of this paper are as follows:
\begin{itemize}
\item We characterized the scale-out behavior of MCM GPU architectures as the number of chiplets increases and identified its limitations.
\item We analyzed the scalability of MCM GPU architectures and its limits under scale-up and hybrid configurations.
\item We evaluated the performance and energy-efficiency implications of both scaling strategies and identified the main factors that constrain their effectiveness.
\item We developed MANGO, an open-source architecture simulator specifically designed for MCM GPUs, enabling the evaluation of novel MCM designs and optimizations.
\end{itemize}

The rest of this paper is organized as follows. Section~\ref{sec:background} provides background on MCM GPUs and  the limitations of prior scalability analyses. Section~\ref{sec:archi_scaling} analyzes scale-out, scale-up, and hybrid scaling strategies for these architectures. Section~\ref{sec:methodology} describes the experimental setup, and methodology, while Section~\ref{sec:simulator} introduces the simulation framework. Section~\ref{sec:evaluation} presents and discusses the experimental results. Section~\ref{sec:related_work} reviews prior work on for chiplet-based architectures. Finally, Section~\ref{sec:conclusion} summarizes the main conclusions.

\section{Background and Motivation}

\label{sec:background} 
\subsection{Multi-Chip-Module GPUs} 

Conventional monolithic GPUs implement all compute units, memory controllers, cache structures, and on-chip interconnects on one large die. As GPU architectures continue to grow, this scaling strategy becomes increasingly difficult to sustain. Larger dies are more expensive to manufacture, more susceptible to defects, and typically suffer from a lower fabrication yield \cite{waldrop2016chips, frank2001device, borkar2011future, hennessy2017computer}. Moreover, increasing the number of compute units within a monolithic GPU can intensify contention over shared resources, such as the crossbar interconnect or the L2 cache, potentially limiting scalability even when additional computational capability is available.

\begin{figure}
    \centering
    \includegraphics[width=\columnwidth]{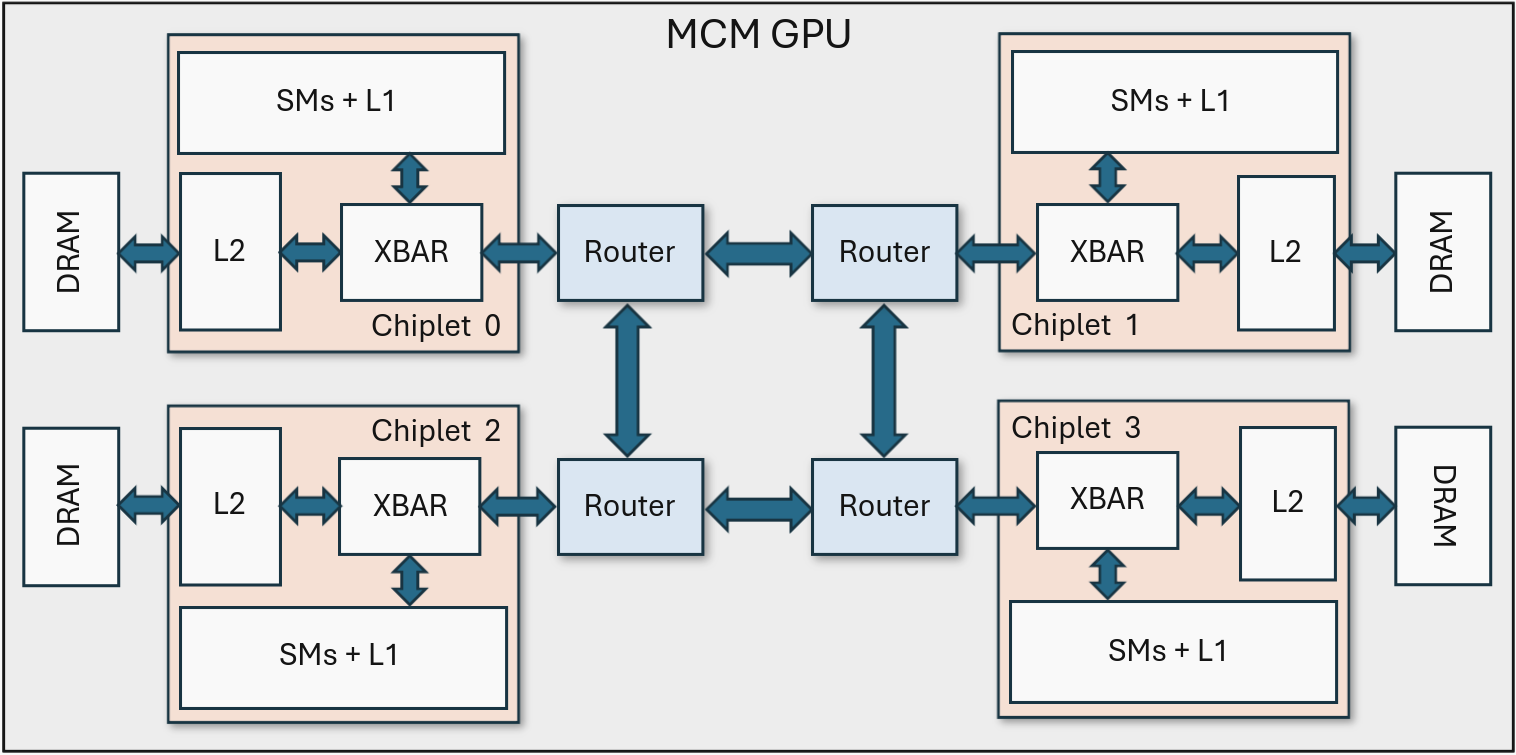}
    \caption{Multi-Chip-Module GPU scheme.}
    \label{fig:MCM}
\end{figure}

The main idea behind MCM GPUs \cite{arunkumar2017mcm, arunkumar2019understanding}, illustrated in Figure~\ref{fig:MCM}, involves partitioning the GPU into multiple chiplets, which can be manufactured independently and then assembled within the same package. Chiplets are organized similarly to monolithic GPUs: each integrates multiple SMs, equipped with a private L1 cache. The L1 caches are connected to the L2 cache through a crossbar. The L2 banks also interface with the memory modules, which reside outside the chiplet.

By enabling the disaggregation of resources across multiple chiplets, MCM-based architectures enable scaling beyond the practical limits of a single die, thereby supporting continued performance growth while mitigating the challenges associated with large monolithic designs \cite{naffziger2021pioneering, arunkumar2017mcm}. This modular organization also improves IP reuse, as the same chiplet architecture can be replicated across different product configurations. Moreover, since smaller dies generally achieve higher manufacturing yield, MCM GPUs can reduce fabrication cost and improve production efficiency. In addition, by relaxing die-area constraints, this approach can provide up to 2$\times$ higher compute density than comparable monolithic GPU designs \cite{arunkumar2017mcm}.

Despite these advantages, MCM GPUs introduce scalability challenges that are not present in monolithic designs. As the number of chiplets and computational resources grow, the memory system and the inter-chiplet interconnect must scale accordingly to sustain the additional parallelism. Since compute units and memory are distributed across multiple chiplets, memory accesses can target data located in a different chiplet \cite{REX, park2025leveraging, wang2026max, zhong2026lrm}. Such remote accesses must traverse the inter-chiplet network to reach the corresponding LLC bank or memory controller, introducing Non-Uniform Memory Access (NUMA) \cite{lameter2013overview} effects, also referred to as Non-Uniform Bandwidth Architecture (NUBA) \cite{nuba} in this context, that increase both latency and network traffic. Therefore, the performance of an MCM GPU depends not only on its aggregate compute capability but also on how efficiently the memory hierarchy and network handle inter-chiplet communication.

\begin{figure*}[t]
    \centering

    \begin{subfigure}[b]{0.49\linewidth}
        \centering
        \includegraphics[]{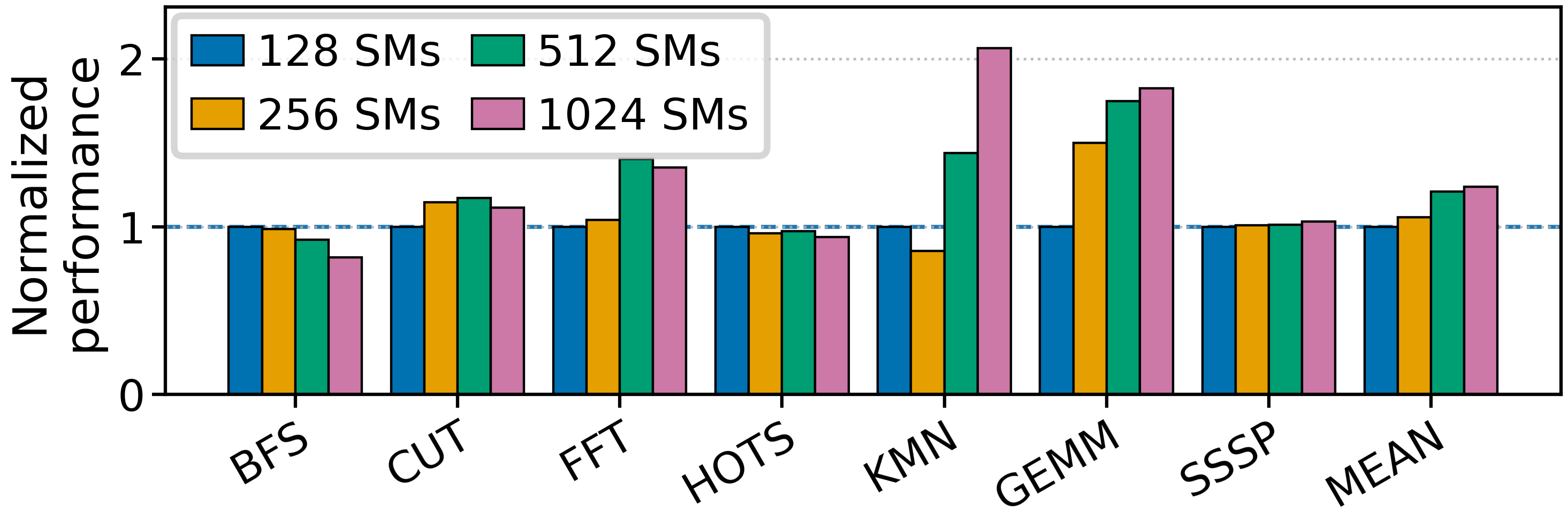}
        \caption{\centering Performance scaling with increasing SM count. \\Normalized to the 128 SMs architecture.}
        \label{fig:sm_scaling_4chiplets}
    \end{subfigure}
    \hfill
    \begin{subfigure}[b]{0.49\linewidth}
        \centering
        \includegraphics[]{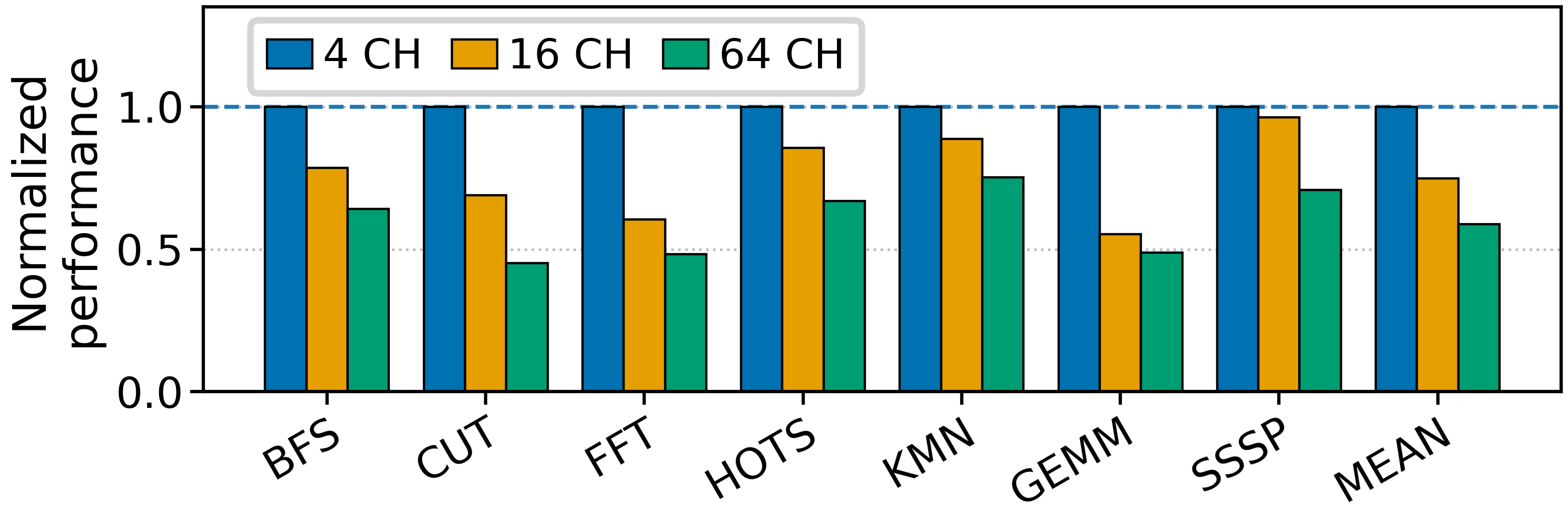}
        \centering\caption{ \centering Performance scaling with increasing chiplet count. \\ Normalized to the 4 chiplet architecture.}
        \label{fig:chiplet_scaling_128sms}
    \end{subfigure}

    \caption{Scale-up and Scale-out strategies}
    \label{fig:motivation}
\end{figure*}

\subsection{The Challenge of Chiplet-Based Scalability} 

Despite its critical role in determining performance, the scalability of MCM GPUs remains largely unexplored. Prior studies on MCM GPUs have focused mainly on architectural mechanisms to reduce the cost of remote accesses \cite{REX, zhang2026location, Data-Sharing-Multi-Chip-GPU}, improving data locality \cite{xu2025exploiting, park2025leveraging, wang2024grit, khairy2020locality}, optimizing inter-chiplet links \cite{li2023ronet, zhang2023seechip}, or designing more efficient communication mechanisms \cite{dalmia2024cpelide, zhong2026lrm}. However, less attention has been paid to understanding how the system behaves when both compute and memory resources are scaled across an increasing number of chiplets. Most existing proposals focus on relatively small and/or fixed-size MCM systems \cite{zhang2026location, zhong2026lrm, park2025leveraging, xu2025exploiting, REX}, commonly assuming four chiplets and a ring network topology. This limits the ability to assess whether the observed trends remain valid for the larger and more disaggregated designs expected from coming GPU generations.

Figure~\ref{fig:motivation} illustrates two complementary scalability limitations of the typical ring-based chiplet architecture. The first experiment, shown in Fig. \ref{fig:sm_scaling_4chiplets}, evaluates \emph{resource scaling within a fixed four chiplet system} (scale-up), by increasing the number of SMs from 128 to 1024, while the associated memory and per-chiplet resources are scaled accordingly. Performance is normalized to the 128-SM configuration. These results show that increasing the per-chiplet compute and memory resources does not necessarily translate into proportional system-level performance. Although available computational resources increase by $8\times$, performance improves by only $1.23\times$ on average at 1024 SMs, corresponding to a scaling efficiency of approximately 15\% relative to linear scaling. Kmeans (KMN) achieves the largest performance, reaching approximately 2.05× in the 1024 SM configuration. However, this gain is caused by a more favorable data placement, which better aligns the data partitions with the execution blocks, improving locality and resource utilization.

The second experiment (\ref{fig:chiplet_scaling_128sms}) evaluates \emph{chiplet-count scaling under a fixed resource budget} (scale-out). Here, the total number of SMs is kept constant at 512, while the system is increasingly disaggregated from 4 to 64 chiplets. Performance is normalized to the four chiplet configuration. Contrarily to the expectation that finer-grained disaggregation could provide greater architectural flexibility, increasing the number of chiplets in a ring network consistently reduces performance across the evaluated applications. On average, performance decreases by approximately $0.76\times$ with 16 chiplets and $0.59\times$ with 64. This performance degradation is caused by increased communication distances, higher inter-chiplet traffic, and greater contention within the ring interconnect.

Together, these results highlight the limitations of both scaling strategies. Scale-up provides limited, application-dependent, performance improvements, while scale-out incurs communication overheads that may outweigh the benefits of disaggregation. Therefore, achieving scalable performance requires jointly considering the amount of resources per chiplet, the degree of disaggregation, and the inter-chiplet network.

Although prior works have evaluated MCM GPUs and chiplet-based accelerators at fixed or multiple chiplet counts, these evaluations are typically tailored to specific architectural mechanisms \cite{arunkumar2017mcm, dalmia2024cpelide, barre, REX} or workload domains \cite{simba2019, AI-MCarquitectures}, rather than applying a general methodology to characterize scalability across different system sizes. In particular, changes to the number of SMs, chiplet count, and interconnect topology have not been studied independently, making it difficult to attribute observed performance trends to additional compute throughput, increased memory-system resources, or growing inter-chiplet communication overhead. 

\section{Architectural Scaling}
\label{sec:archi_scaling}


The previous limitations leave three fundamental open questions. First, it remains unclear whether increasing compute resources leads to proportional improvements in end-to-end performance and energy efficiency. Second, the risk of the inter-chiplet interconnect becoming the dominant system bottleneck under increased remote-memory traffic, communication latency, and network contention has not been assessed. Third, the impact of different scaling strategies on energy consumption and efficiency remains poorly understood. Answering these questions is essential not only for understanding the scalability limits of current MCM GPU architectures, but also for providing industry with actionable design guidelines of future MCM GPU systems, which may vary depending on the requirements of each industry segment.

\subsection{Scale-out}

This strategy provides a high degree of modularity, as the same chiplet design may be reused across products with different performance, power, and cost targets. As the architecture of the chiplet remains fixed, its power consumption, manufacturing cost, and performance can be characterized in advance. Consequently, scale-out reduces design and verification effort and facilitates the development of a family of products based on a common reusable building block.

However, the main limitation of this approach is the inter-chiplet network. As the number of chiplets increases, a larger fraction of the memory requests need to traverse the network, while the average communication distance and the contention over the communication links increase. This may prevent additional compute resources from being effectively utilized. Such effect is particularly visible for ring-based architectures as displayed in Fig.~\ref{fig:chiplet_scaling_128sms}, which shows performance degradation despite a linear growth in compute capability, caused by the inter-chiplet network becoming a scalability bottleneck. The average communication latency increases by $3.84\times$ and $16.22\times$ with 16 and 64 chiplets respectively, compared to the 4 chiplet configuration. This sharp increase reflects the longer communication distances and growing contention that prevent the additional compute resources from being fully utilized.

Workload distribution and data placement further influence this behavior. An unfavorable placement can concentrate remote memory requests on a limited set of destinations and network links, as observed for Kmeans in Fig.~\ref{fig:sm_scaling_4chiplets}, where the 1024 SM configuration benefits from better alignment between data and execution blocks. These adverse traffic patterns create communication hotspots, increase queuing delays, and can saturate the  network before the available computational resources are fully utilized. NUMA effects therefore become increasingly significant as the system grows, since accesses to remote memory modules consume more network resources and experience higher latency than local accesses.

\subsection{Scale-up}

The scale-up strategy is based on provisioning additional resources within each chiplet while keeping the number of chiplets constant. This primarily involves increasing the number of SMs, together with the cache capacity, memory banks, schedulers, and bandwidth required to sustain their execution. Scale-up therefore represents the most direct strategy for increasing the computational capability of the system by concentrating additional resources within a larger and more capable chiplet, rather than distributing them.

\begin{figure}
    \centering
    \includegraphics[]{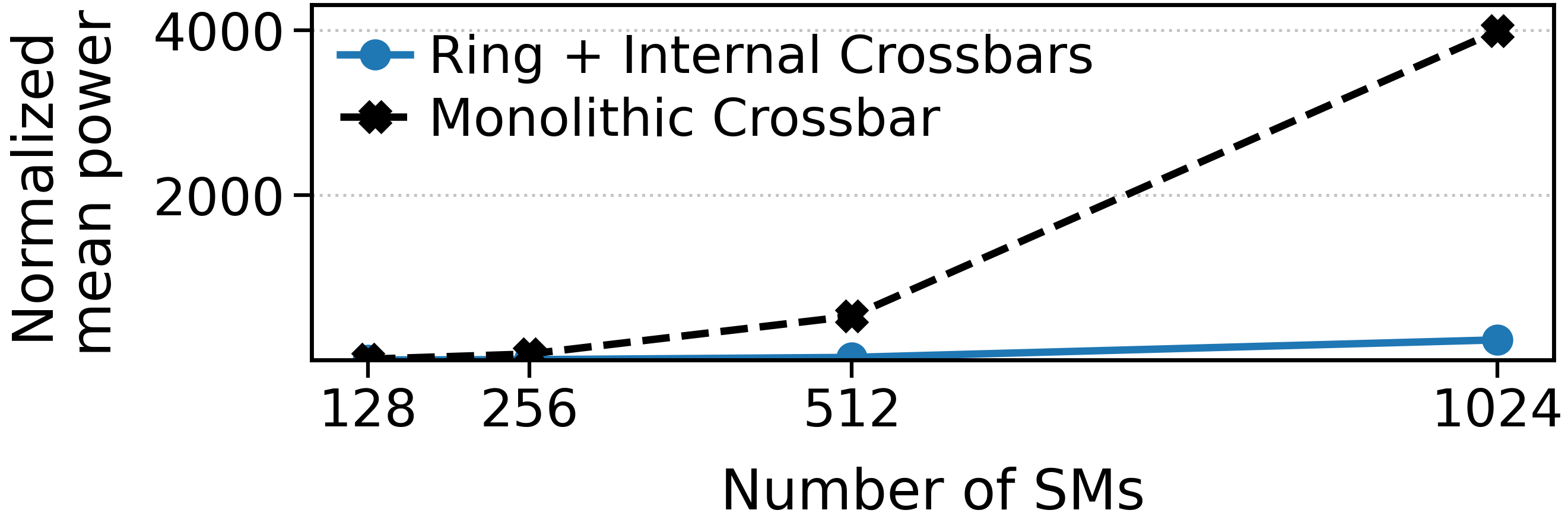}
    \caption{\centering Power scaling relative to a 128 SM, 4 chiplet ring.}
    \label{fig:sm_scaling_ring_monolithic}
\end{figure}

However, increasing the resources within each chiplet also enlarges the internal structures required to connect and coordinate them. In particular, crossbars and shared datapaths must support more endpoints, ports, and concurrent accesses. Since the area and wiring complexity of a crossbar can grow quadratically with its number of ports, its power consumption increases rapidly as the architecture scales. Figure ~\ref{fig:sm_scaling_ring_monolithic} shows how the power for the monolithic design grows sharply with the number of SMs, while the ring based architecture shows a much more moderate increase, by distributing communication across simpler interconnect structures.

Increasing chiplet size also entails a significant economic cost. A larger die occupies more area, reducing the number of chiplets that can be manufactured from a single wafer. Because manufacturing defects are distributed across the wafer, a larger chiplet has a higher probability of containing at least one defect. These costs are further amplified by the additional engineering effort required to design and manufacture a large chiplet. They may also require more advanced packaging, stronger power delivery, and more expensive cooling solutions to manage their higher total power and thermal density.

Scale-up therefore exposes a fundamental trade-off. Increasing the resources within each chiplet can preserve architectural balance and provide stable performance scaling, but it also increases internal interconnection overhead, power consumption, design complexity, and manufacturing cost. Beyond a certain chiplet size, the marginal performance benefit of adding resources may no longer compensate for the growing implementation and economic penalties.

\subsection{Hybrid scaling}

Between the two extreme strategies of scale-out and scale-up, lies a broad range of intermediate configurations, referred to here as \textit{hybrid scaling}. This approach combines moderately more capable individual chiplets with the coordinated deployment of multiple chiplets. By balancing both dimensions, hybrid scaling addresses the limitations identified for the two extreme strategies. As a result, it limits the growth of internal crossbars and shared datapaths within each chiplet while reducing the number of chiplets that must communicate through the inter chiplet network. This balance mitigates the area and energy costs of scale-up and the bandwidth, latency, and topology requirements of scale-out.

This mixed strategy provides greater flexibility and enables the system configuration to be adapted to the specific requirements of each workload. Therefore, the different hybrid scaling configurations must be evaluated to identify the most appropriate balance between system capacity, resource utilization, complexity, and performance.

However, identifying an effective hybrid scaling configuration is substantially more complex than evaluating either extreme in isolation, as the design space grows rapidly and requires a systematic analysis of a large number of feasible combinations. The interaction between scale-up and scale-out is often non-linear, meaning that improvements in one dimension may introduce bottlenecks or inefficiencies in another, reflecting workload-specific behavior or measurement noise rather than a generally effective balance.

\subsection{Inter-chiplet Network}

The inter-chiplet network plays an important but different role depending on the selected scaling strategy. Its latency, bandwidth, topology, and ability to handle contention determine the cost of transferring data and coordinating computation across chiplets. Consequently, the benefits obtained by increasing the number or capability of the chiplets cannot be evaluated independently of the communication fabric.

\begin{figure}
    \centering
    \includegraphics[width=0.9\linewidth]{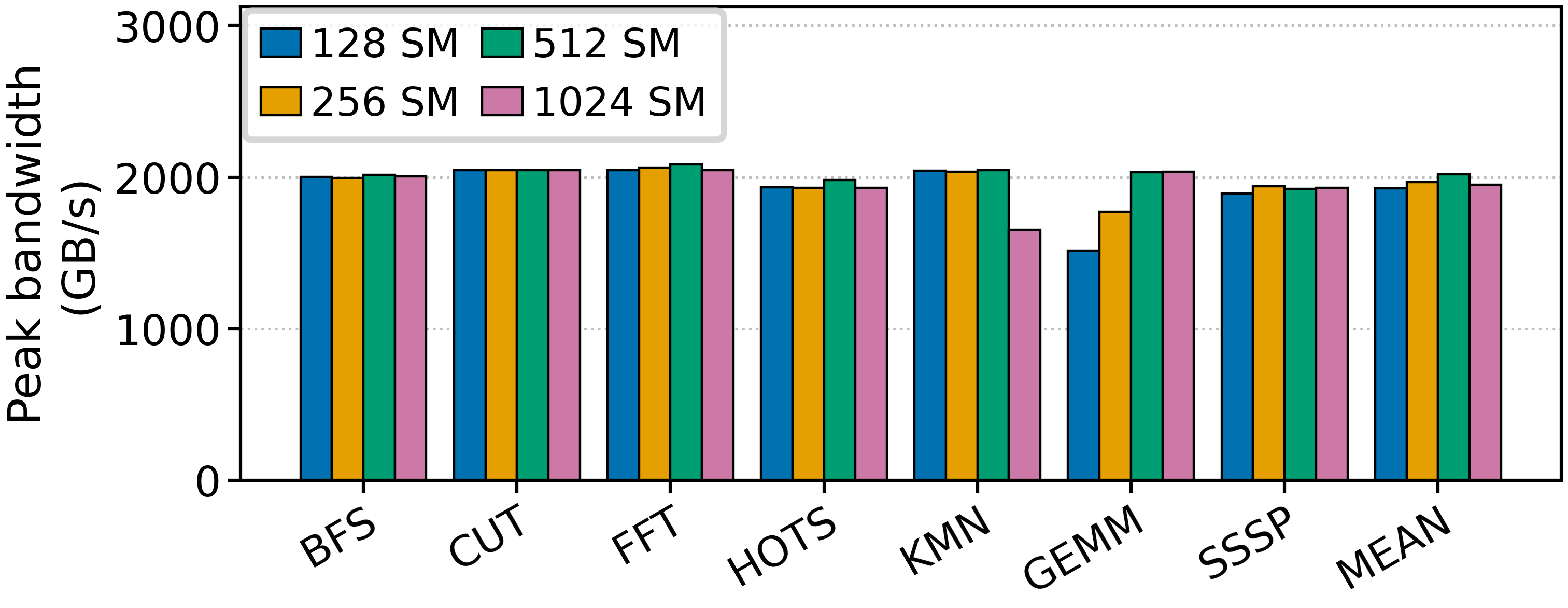}
    \caption{\centering Peak Bandwidth scaling with increasing SM counts in a four chiplet ring design.}
    \label{fig:band_4_chiplets}
\end{figure}

In a \textit{scale-up} configuration, resources are concentrated in a smaller number of chiplets. This reduces the number of communicating endpoints and can limit communication, synchronization, and routing overheads. However, each chiplet may generate a larger volume of traffic, placing significant pressure on the network interfaces and on the network fabric itself. As shown in Figure~\ref{fig:band_4_chiplets}, increasing the total number of SMs from 128 to 1024 does not produce a proportional increase in the peak bandwidth of the four chiplet ring. For most workloads, peak bandwidth remains close to peak of the network (2048 GB/s), showing that the ring topology is already saturated at 128 SMs. GEMM is the only exception, reaching approximately $0.75\times$ and $0.86\times$ the peak bandwidth for the 128 and 256 SMs configurations, respectively, indicating that the ring is not saturated with 128 SMs and can accommodate the increased communication associated with more SMs per chiplet.
Kmeans, in contrast, exhibits a substantial bandwidth reduction in the 1024 SM configuration. This is the result of a more favorable alignment between memory accesses and page placement, which reduces remote traffic.

In contrast, \textit{scale-out} configurations distribute resources across a larger number of smaller chiplets. Although the traffic generated by each individual chiplet may be lower, the number of communication flows, routing decisions, and synchronization operations increases. Data may also need to traverse additional network hops, increasing latency and energy consumption. As a result, scale-out is more sensitive to network topology, routing policy, link contention, and communication locality. In particular, topologies providing greater path diversity and higher bisection bandwidth can distribute concurrent communication flows more effectively, reducing congestion, and improving network utilization. 

If the workload requires frequent data exchanges across chiplets, network overhead can compromise performance. This behavior is evident in Figure~\ref{fig:chiplet_scaling_128sms} for the ring topology, as introduced before. In addition, all evaluated benchmarks experience performance degradation as the degree of disaggregation increases. CUTCP represents the most extreme case: when scaling to 64 chiplets, its normalized performance falls to approximately 0.45, indicating that more than half of its baseline performance is lost due to the increased disaggregation.

The impact of the network on \textit{hybrid scaling} configurations lies between these two extremes. By combining a moderate number of chiplets with moderate per-chiplet capability, this strategy can reduce both the traffic concentration in scale-up and the communication overhead associated with scale-out. Nevertheless, the optimal design depends on how the workload maps onto the chiplets, the amount and locality of the traffic, and the network's ability to sustain concurrent communication.

\section{Methodology}
\label{sec:methodology}

The evaluation conducted in this paper compares different MCM GPU configurations under realistic workloads. For each configuration, we collect system-, energy- and network-level metrics to identify the factors that limit scalability.

\subsection{Architectural Configurations}

We evaluated several MCM GPU configurations with different numbers of chiplets and compute resources. Depending on the objective of each experiment, these parameters are varied independently or jointly. 

For \textit{scale-out}, the evaluated systems contain 4, 16, 36 or 64 GPU chiplets, organized as regular (2$\times$2), (4$\times$4), (6$\times$6), and (8$\times$8) layouts, respectively. A 36 chiplet configuration is considered instead of a 32-chiplet design to preserve a regular square and a network with edge-symmetry. 
In these experiments, each chiplet will be composed by 16 SMs per chiplet and, following the scaling methodology adopted in \cite{mallya2025performance}, the number of memory controllers and L2 banks increases linearly with the number of GPU chiplets. Thus, the evaluated configurations maintain a one-to-one ratio between compute chiplets and memory dies.

The \textit{scale-up} and \textit{hybrid scaling} studies evaluate four system sizes, with a total of 128, 256, 512, and 1024 SMs. For each size, the compute resources are partitioned into 4, 16, or 64 chiplets, enabling configurations with the same aggregate computational capability but different architectures to be compared. This experimental design allows the scalability to be characterized along the two complementary dimensions: increasing the total number of SMs and distributing a fixed number of SMs across a larger number of chiplets.

Table~\ref{tab:gpgpusim_config} summarizes the main architectural and memory-system parameters of the chiplet configuration considered in this work. This configuration was selected because it closely resembles contemporary GPU architectures, making it a representative baseline for our evaluation.  The total L2 cache capacity is scaled according to the number of SMs in each evaluated configuration, thereby maintaining a comparable amount of cache capacity per SM.

\begin{table}[t]
    \centering

    \scriptsize
    \setlength{\tabcolsep}{2pt}
    \begin{tabular}{|p{1.65cm}|p{6.75cm}|}
    \hline
\textbf{Parameters} & \textbf{Value} \\

\hline
SM & 1.905GHz, 64K regs., 1K threads, 32 warps, 4 warp schedulers\\
\hline
Execution Units & 4 SP, 4 INT, 4 SFU and 4 Tensor units per SM \\

\hline 
Block Scheduler & Round-Robin across chiplets \\ 

\hline
Shared Memory & 64KiB/SM, 32 banks, 20-cycle latency \\

\hline
L1 Data Cache & 64KiB/SM, 4 banks, 128B lines ($4\times$32B sectors), 20-cycle latency \\

\hline
L2 Cache & 128KiB/SM, 2 banks, same lines as L1, 160-cycle latency \\

\hline
Memory System & GDDR6-3.5GHz, latency 100 cycles, 2 sub-partitions per channel \\



\hline

    \end{tabular}
        \caption{Overall chiplet configuration.}
    \label{tab:gpgpusim_config}
\end{table}

\subsection{Workloads and metrics}

The workloads used in this study are summarized in Table~\ref{tab:benchmarks}. They were selected from the Rodinia \cite{che2009rodinia} and Parboil \cite{parboil} benchmark suites, and the FNNDSC repository \cite{fnddsc_dijkstra_kernel} to cover diverse computate and communication behaviors. The selected set includes irregular graph workloads such as BFS and SSSP, the compute-intensive GEMM kernel, iterative Kmeans, a regular stencil-based Hotspot benchmark, the structured communication pattern of FFT, and the three-dimensional computation performed by Cutcp. Each benchmark is limited to a fixed number of executed instructions in all scenarios. This guarantees results remain comparable across experiments, as all design points perform the same amount of work, allowing energy consumption and performance differences to be attributed directly to communication overheads.

The datasets were selected to provide sufficiently large and representative executions, achieving an average GPU occupancy above 95\% during the measured execution interval. Maintaining high occupancy ensures that most of the available processing resources are active, avoiding artificially underutilized conditions and exposing the inter-chiplet network to sustained computation and communication activity. Evaluation follows a strong-scaling methodology: the benchmark inputs and kernel launch parameters remain unchanged across all scaling configurations, while only the amount and organization of the available hardware resources vary. Consequently, each configuration executes the same benchmarks and the same number of instructions, allowing a direct assessment of how scaling affects performance and energy. 

\begin{table}[t]
    \centering
    \scriptsize
    \setlength{\tabcolsep}{2pt}
    
    \begin{tabular}{
        |>{\centering\arraybackslash}m{1.5cm}
        |>{\centering\arraybackslash}m{1.5cm}
        |>{\centering\arraybackslash}m{2.75cm}
        |>{\centering\arraybackslash}m{2cm}|
    }
        \hline
        \textbf{Benchmark} &
        \textbf{Suite} &
        \textbf{Dataset} &
        \textbf{Kernel size (Grid), (Block)} \\
        \hline

        BFS & Rodinia & soc-Pokec (SNAP) & (3190,1), 
                                           (512,1)\\
        \hline
        GEMM & Parboil & matrix2 (Parboil) & (64,128), (16,8) \\
        \hline
        Kmeans & Rodinia & kdd\_cup (Rodinia) & (4096,1), (256,1)\\
        \hline
        Hotspot & Rodinia & temp\_1024 (Rodinia) & (86,86), (16,16) \\
        \hline
        SSSP & FNNDSC & soc-LiveJournal1 (SNAP) & (269507,1), (256,1)\\
        \hline
        FFT & Parboil & N/A & (1024,1), (1024,1) \\
        \hline
        Cutcp & Parboil & watbox.sl100.pqr (Parboil) & (77, 102, 1), (8,2,8) \\
        \hline
        
    \end{tabular}
    \caption{Benchmarks.}
    \label{tab:benchmarks}
\end{table}

For each benchmark, four main performance and energy metrics are collected: instructions per cycle (IPC), energy consumption, energy efficiency, and peak bandwidth. Energy consumption includes both compute and communication activity. The network energy is calculated by multiplying the average network power by the corresponding execution time. In the evaluated MCM GPU systems, the communication energy is divided into two components: the energy consumed by the internal crossbars within each chiplet and the energy consumed by the inter-chiplet network. Energy efficiency complements the previous metric by relating computational performance to energy consumption, expressed as IPC per joule.

\subsection{Inter-chiplet Network Configurations}
\label{subsec:net-confi} 

\begin{figure}[t] 
    \centering
    
    \begin{subfigure}[t!]{0.20\textwidth}
        \centering
        \includegraphics[scale=0.12]{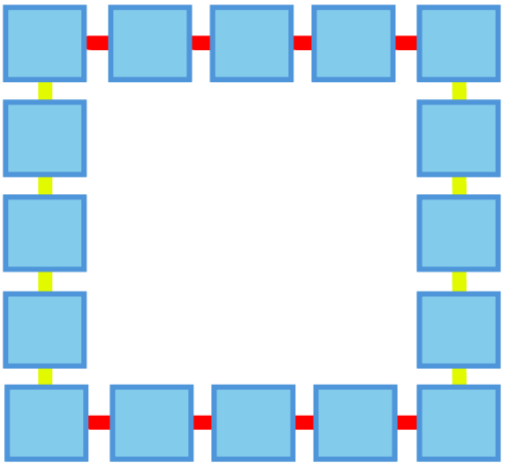}
       \caption{Ring}
        \label{fig:ring}
    \end{subfigure}
    \begin{subfigure}[t!]{0.20\textwidth}
        \centering
        \includegraphics[scale=0.10]{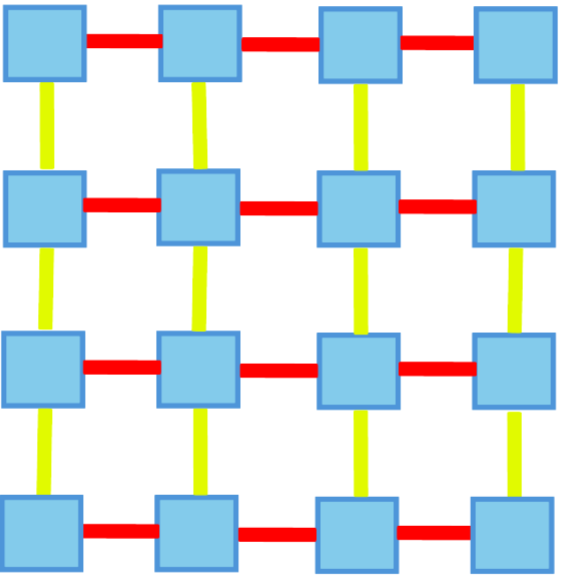}
        \caption{Mesh}
        \label{fig:mesh}
    \end{subfigure}
    
    \begin{subfigure}[t!]{0.20\textwidth}
        \centering
        \includegraphics[scale=0.10]{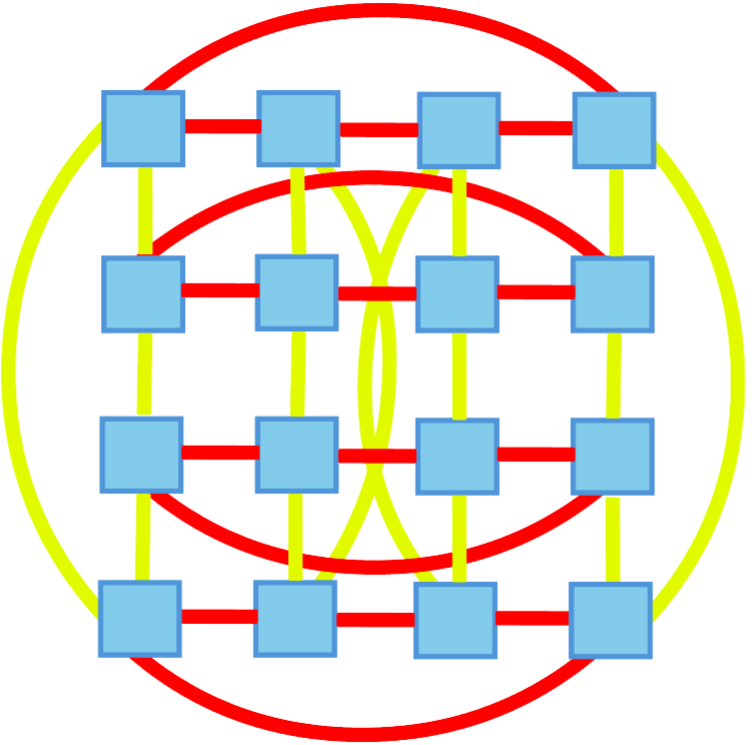}
        \caption{2D Torus}
        \label{fig:2D Torus}
    \end{subfigure}
    \begin{subfigure}[t!]{0.20\textwidth}
        \centering
        \includegraphics[scale=0.10]{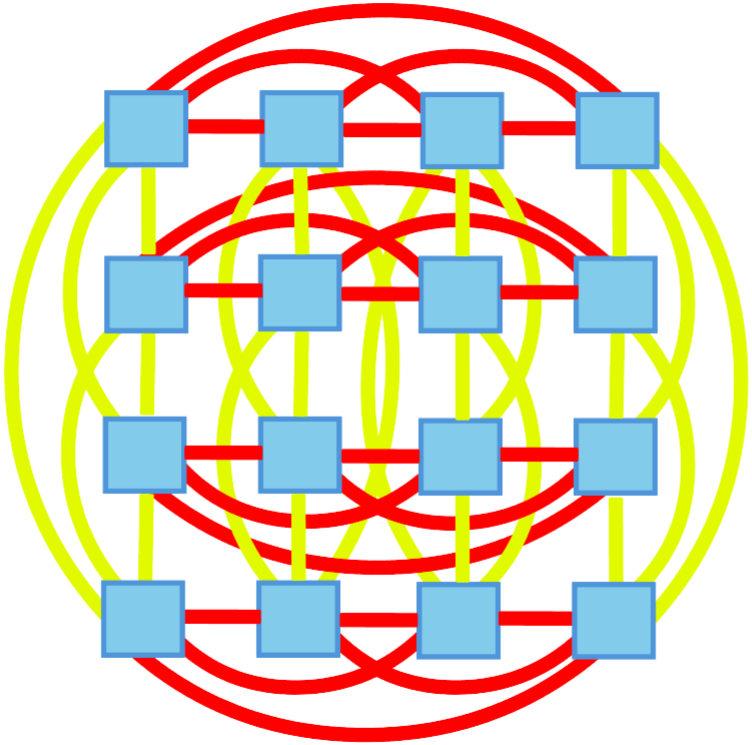}
        \caption{Flattened Butterfly}
        \label{fig:butterfly}
    \end{subfigure}
    
    \caption{\centering Evaluated topologies for a 16 chiplet network. Red lines: horizontal links; Yellow lines: vertical links.}
    \label{fig:inter-chiplet}
\end{figure}

Four network topologies, which can be seen in Fig.\ref{fig:inter-chiplet}, are evaluated: Ring, and 2D versions of Torus (2D-Torus), Mesh, and Flattened Butterfly (Flat-B). Both performance and cost directly depend on the number of router ports used by each topology; note that the number of links of the whole network is exactly half the number of total router ports. But performance also inversely depends on the average distance between routers. Assuming random traffic in a symmetric network, its maximum achievable throughput is limited by the total number of network ports divided by the average distance. This average distance also dictates the standard packet latency, whereas the diameter bounds its maximum value.

The bidirectional ring is a very cheap solution, since it requires only two network ports, or one link, per router, but its diameter and average distance increase linearly with the size of the network. Therefore, in systems that handle high volumes of traffic, its maximum size must be quite small to avoid saturating the network. The ring is symmetric, which is highly desirable; however, it is prone to deadlock due to the cyclic nature of its topology. It is  known that the deadlock avoidance mechanism used seriously affects performance. Most mechanisms employ several virtual channels per port, which increase both cost in the router area and energy consumption. Routing is straightforward, as the source only needs to decide in which direction to send the packet. 

To overcome the scaling issues of the ring, a 2D torus can be used. It requires routers with 4 network ports. The diameter and average distance decrease quadratically compared to the ring, allowing for networks with enough capacity to serve dozens of endpoints. Square 2D tori are fully symmetric, but, like rings, they require specific deadlock-avoidance mechanisms. Dimensionally Ordered Routing (DOR) is a typical deterministic routing solution for tori, which are deadlock-free when their rings are visited in dimension order and they feature mechanisms to make them individually deadlock-free. Similarly to ring, performance is highly dependent on such mechanisms. Adaptive routing typically improves performance, but at a higher cost.

A 2D mesh is a torus from which the wraparound links have been removed. This causes it to lose symmetry and increases the distances between pairs of routers, leading to lower performance. Nevertheless, it has at least three advantages. First, it is completely deadlock-free when using DOR, without requiring additional resources, such as those needed by the rings that make up tori. Second, it frees up ports on peripheral routers to connect to other functional units. And finally, it can use short links between routers whose length does not increase with the network size. Regarding this, a folded torus prevents the growth of its wraparound links, but the length of the resulting links doubles compared to the mesh.

A square 2D Flattened Butterfly is a symmetric network, in which the rows and columns are complete graphs. That is, there is a link between each pair of routers in each row or column. Hence, its routers require as many ports as the number of routers per dimension minus one, multiplied by 2. The network diameter is only 2, since at most one hop per dimension is required with minimal routing; its average distance is just below 2. When using DOR, it is deadlock-free, without need for additional resources, just like a mesh. However, this topology requires a large number of links of varying lengths, with the longest ones connecting the most distant routers within each row and column. To ease the implementation of such links and mitigate their cost, concentration schemes may be used, based on increasing the chiplet to router ratio. For example, for a 64 chiplet system, 4 chiplets may be connected to each router, resulting in a $4 \times 4$ Flattened Butterfly (Flat-B) with 16 routers and substantially fewer network links.

The physical inter-chiplet interface has been modeled in our simulations according to UCIe 3.0~\cite{ucie3.0}. Each unidirectional channel operates at 2~GHz with 256-B flits, yielding an ideal raw bandwidth of $2\times10^{9}\times256=512~GB/s$. This matches a 64-lane UCIe interface operating at 64~GT/s, implemented as either one x64 Advanced Package module for links up to 2~mm or four x16 Standard Package modules for links up to 25~mm. The network model captures finite buffering, credit-based flow control, backpressure, and contention, which are the main mechanisms governing traffic-dependent performance.
     
\section{Simulation Environment}
\label{sec:simulator} 


Only a limited number of public frameworks can model MCM GPUs or closely related multi-chiplet GPU systems. Among the most relevant are MGPUSim\cite{MGPUSim}, LEGOSim\cite{legosim}, and the multi-chiplet GPU gem5 extension used by CPElide\cite{dalmia2024cpelide}. These frameworks approach MCM simulation from different architectural and network-modeling perspectives, as summarized in Table~\ref{tab:mcm_gpu_frameworks}.

\begin{table}[t]
    \centering
    \scriptsize
    \setlength{\tabcolsep}{1pt}
    \renewcommand{\arraystretch}{1}

   \begin{tabular}{
        |>{\centering\arraybackslash}m{1.75cm}
        |>{\centering\arraybackslash}m{1.2cm}
        |>{\centering\arraybackslash}m{1.4cm}
        |>{\centering\arraybackslash}m{0.9cm}
        |>{\centering\arraybackslash}m{0.8cm}
        |>{\centering\arraybackslash}m{2.2cm}|
    }
        \hline
        \textbf{Framework} &
        \textbf{GPU Model} &
        \textbf{Interconnect} &
        \textbf{App. Changes} &
        \textbf{Energy Model} &
        \textbf{Main Considerations} \\
        \hline

        MGPUSim \cite{MGPUSim} &
        AMD GCN &
        PCIe-like tree &
        No &
        No &
        System-level multi-GPU execution. \\
        \hline
        gem5-CPElide \cite{dalmia2024cpelide} &
        AMD-like &
        Crossbar & 
        No &
        Partial &
        Inter-chiplet coherence and synch. \\
        \hline
        LEGOSim \cite{legosim} &
        Hetero-geneous &
        Message passing &
        Yes &
        Yes &
        Explicit communication only. \\
        \hline
        \textbf{MANGO} &
        NVIDIA-like &
        Configurable BookSim network &
        No &
        Yes &
        Detailed intra- and inter-chiplet network. \\
        \hline
    \end{tabular}
    \caption{Summary of MCM GPU simulation frameworks.}
    \label{tab:mcm_gpu_frameworks}
\end{table}

MGPUSim is a cycle-accurate simulator for AMD GCN-based multi-GPU systems, which supports multiple interconnected GPU devices and remote memory accesses. However, MGPUSim's interconnection is based on a PCIe-oriented tree network and abstracts several router-level details. Therefore, it offers limited support for systematically exploring topologies, routing algorithms or channels. Moreover, its AMD GCN execution model limits direct comparison with previous MCM GPU proposals\cite{arunkumar2017mcm,khairy2020locality,zhang2026location,zhong2026lrm,REX,Data-Sharing-Multi-Chip-GPU}, which mainly use NVIDIA architectures. In particular, the original MCM GPU \cite{arunkumar2017mcm} proposal relies on specific mechanisms for its chiplet organization, including distributed block scheduling, remote data caching, and first touch page placement. Reproducing these mechanisms in MGPUSim would require adapting its execution and memory system models, complicating a direct comparison under equivalent architectural assumptions.

The gem5 multi-chiplet GPU extension, used to evaluate CPElide, models an AMD-like GPU with multiple chiplets. However, its published configuration primarily targets inter-chiplet synchronization and coherence, and uses a crossbar interconnection, rather than a configurable router-based network. Additionally, its dependence on an older ROCm software stack introduces software limitations that restrict the maximum number of chiplets to seven.

LEGOSim provides a framework for heterogeneous multi-chiplet systems by integrating separate simulators through a common communication interface. Inter-chiplet communication must be explicitly exposed through message-passing and data-transfer operations, requiring modifications to the workloads or their simulation wrappers, which results in an unrealistic programming model for MCM architectures. Furthermore, LEGOSim models communication between independent simulated components rather than fine-grained memory traffic generated during execution within a single logical MCM GPU.

To address these limitations, this work proposes MANGO (\textbf{M}ulti-chiplet \textbf{A}rchitecture and \textbf{N}etwork simulator for \textbf{G}PU \textbf{O}rganization)\footnote{The MANGO source code and documentation will be made publicly available on GitHub upon acceptance of the paper.}, an extension of GPGPU-Sim that incorporates BookSim-based modeling of the inter-chiplet interconnection. GPGPU-Sim \cite{khairy2020accel} provides an established NVIDIA-like execution model and supports the execution of unmodified CUDA applications, thus preserving the original programming model. It models workload execution, SIMT cores and the memory hierarchy. BookSim \cite{jiang2010booksim} provides detailed and configurable models of both intra-chiplet and inter-chiplet networks, including routers, routing algorithms, buffers, and network topologies. By integrating both tools, the resulting framework captures communication within and across chiplets without explicit message-passing operations or modifications to simulated application.

In addition, our simulation framework provides a compatible and fully integrated power model through AccelWattch~\cite{khairy2020accel}. This compatibility allows to evaluate both performance and energy consumption using the same simulated GPU configuration. The energy consumed by the networks is estimated using the BookSim power model, which analyzes the dynamic, clock, and leakage power of the channels, buffers, crossbars, and output circuitry. The configuration of the energy parameters is based on the UCIE 3.0 standard \cite{ucie3.0} and \cite{shabani2023hierarchical, energia-hpca-chiplet}. 

\section{Evaluation}
\label{sec:evaluation}

This section addresses the following open questions:

\begin{itemize}
    \item Given a fixed-size chiplet (SMs and memory resources), does the addition of more chiplets to the GPU lead to sustained scaling of performance?
    \item What is the impact of disaggregation on performance across different degrees of scale?
    \item What is the impact of chiplet-based scaling on energy consumption and efficiency?
\end{itemize}

\subsection{Scale-out}

\begin{figure}[t] 
    \centering
    
    \begin{subfigure}[b]{0.49\columnwidth}
        \centering 
        \hspace*{-0.03\textwidth}\includegraphics[width=1.03\textwidth]{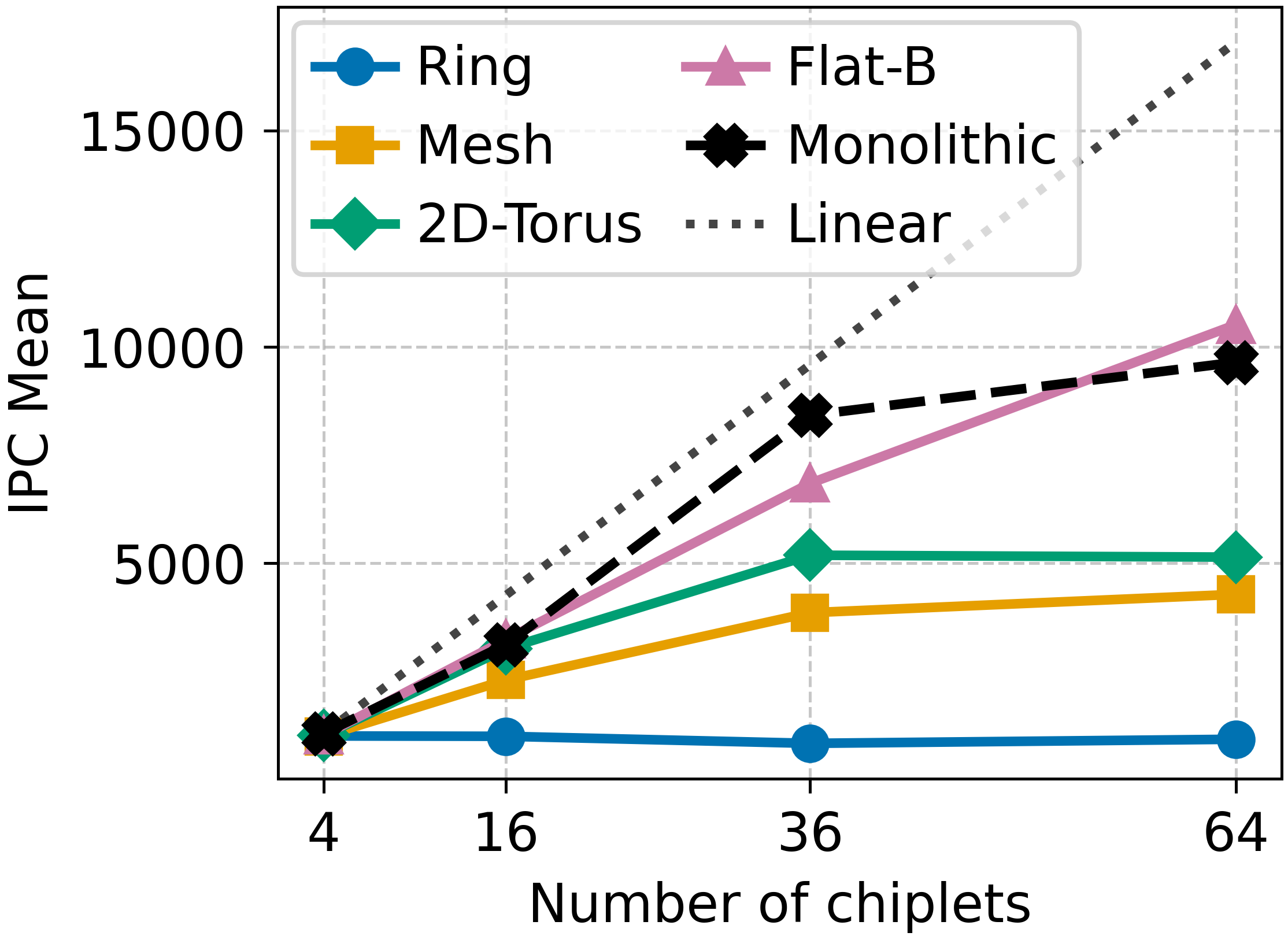}
       \caption{Average.}
        \label{fig:sub_ipc_growing_svg}
    \end{subfigure}
    \hfill
    \begin{subfigure}[b]{0.49\columnwidth}
        \centering
        \includegraphics[width=\textwidth]{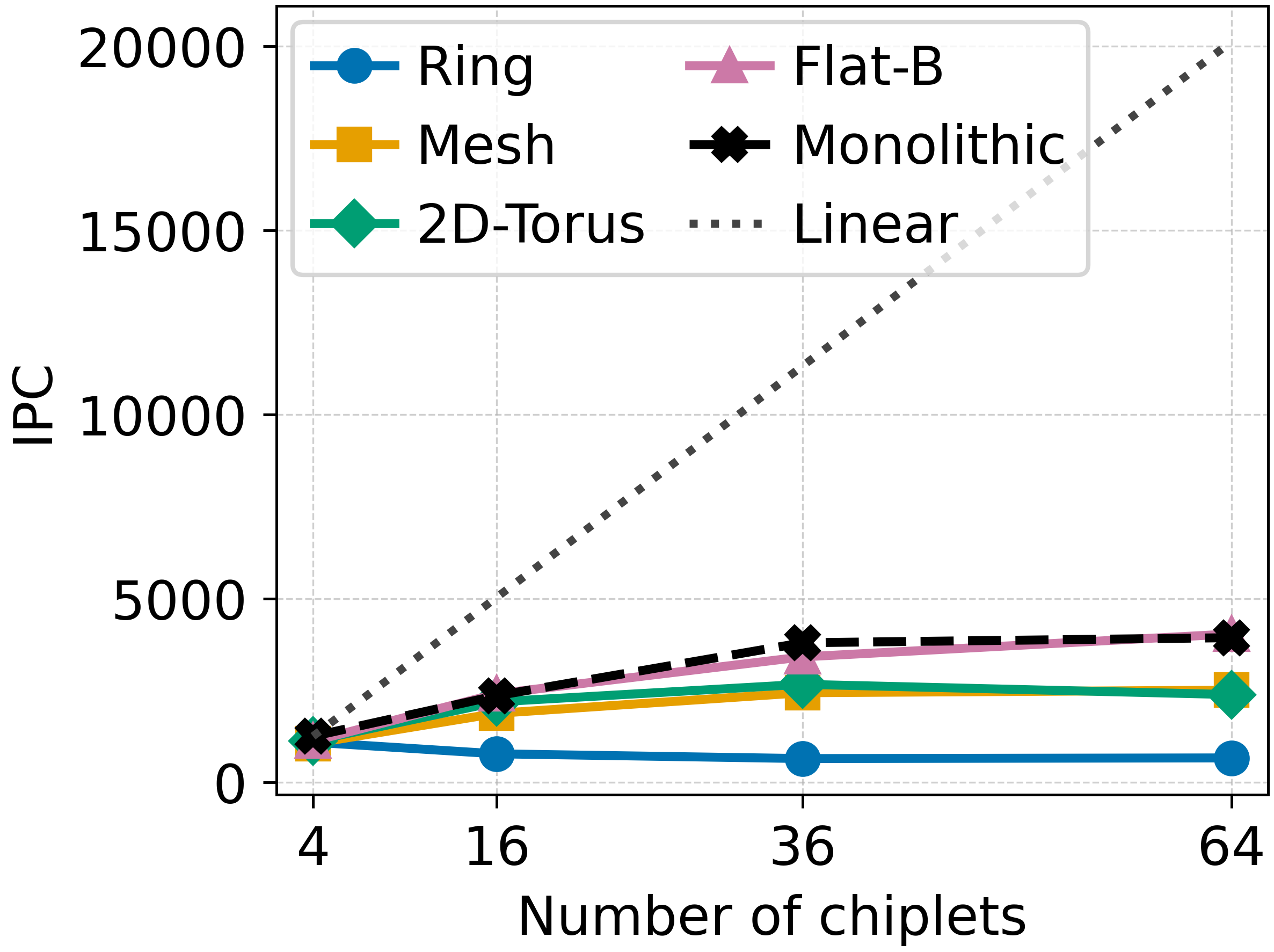}
        \caption{BFS.}
        \label{fig:sub_ipc_growing_bfs}
    \end{subfigure}
    
    \vspace{0.5em}
    
    \begin{subfigure}[b]{0.49\columnwidth}
        \centering
        \includegraphics[width=\textwidth]{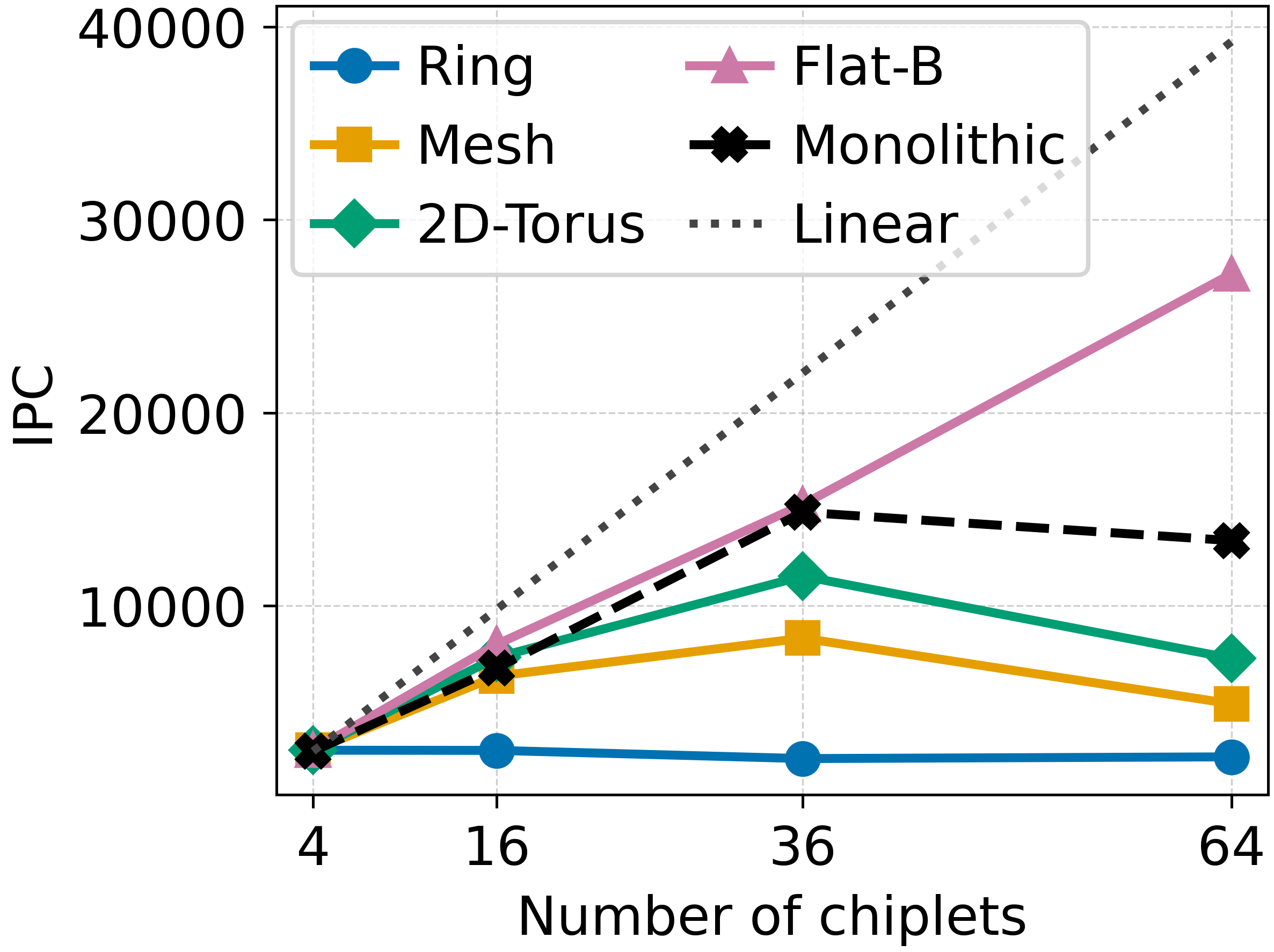}
        \caption{FFT.}
        \label{fig:sub_ipc_growing_fft}
    \end{subfigure}
    \hfill
    \begin{subfigure}[b]{0.49\columnwidth}
        \centering
        \includegraphics[width=\textwidth]{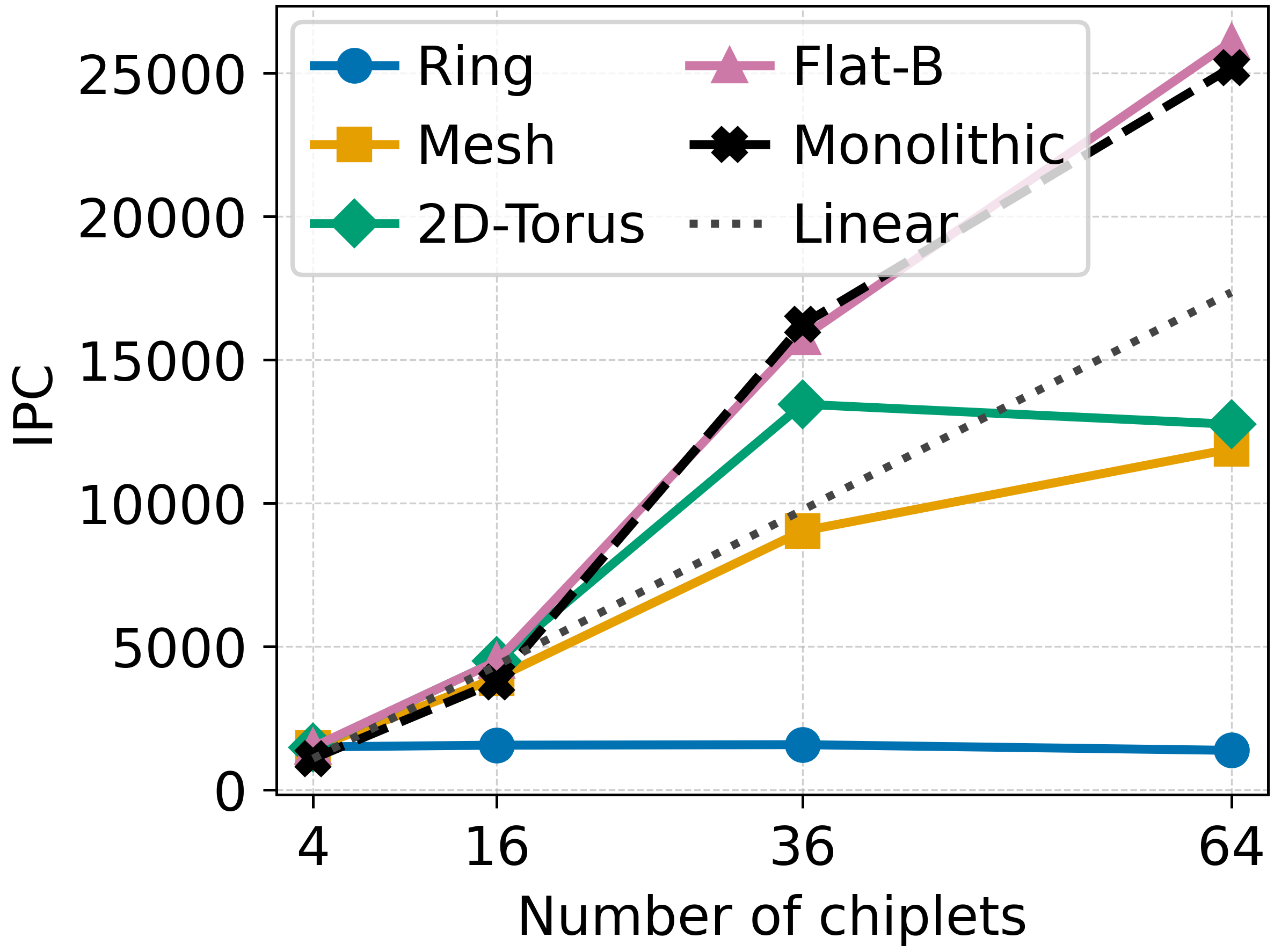}
        \caption{GEMM.}
        \label{fig:sub_ipc_growing_sgemm}
    \end{subfigure}
    \caption{\centering IPC per topology with growing number of chiplets.}
    \label{fig:fixed_sm_size_avg}
\end{figure}

Figure \ref{fig:fixed_sm_size_avg} shows the evolution of IPC for a growing number of chiplets, each with a fixed size of 16 SMs and the configuration shown in Table \ref{tab:gpgpusim_config}. Consequently, with each additional chiplet the GPU grows in both potential compute capability and memory throughput. The figure compares the four interconnection topologies explained in Section \ref{subsec:net-confi} for configurations ranging from 4 to 64 chiplets. The performance for an equivalent monolithic GPU and a linear scaling curve are also included as a reference. The latter represents the IPC that would be achieved under perfectly linear scaling, projected from the IPC of the smallest monolithic configuration.

Figure \ref{fig:sub_ipc_growing_svg} reports the average results across all evaluated applications. Figures \ref{fig:sub_ipc_growing_bfs}, \ref{fig:sub_ipc_growing_fft} and \ref{fig:sub_ipc_growing_sgemm} present the results for BFS, FFT, and GEMM, respectively, highlighting their specific behaviors. The trends for the applications not shown closely resemble those of either GEMM, FFT or BFS. 

As discussed in the motivation for this work, Ring topologies, the current industry standard for MCM-based GPUs, fail to scale beyond very small chiplet counts. As shown in Figure \ref{fig:sub_ipc_growing_svg}, the IPC for the Ring plateaus or even decreases as the number of chiplets grows. This effectively eliminates the benefits of scale-out beyond a small number of chiplets. 

The remaining configurations offer different levels of scalability, with the importance of topology growing with system size. At the 16 chiplet design point, the Mesh, 2D Torus and Flat-B deliver comparable performance. Differences become apparent at 36 chiplets, where Flat-B achieves approximately $1.32\times$ and $1.77\times$ the performance of the Torus and Mesh, respectively. However, both topologies exhibit diminishing returns at larger networks, with performance barely increasing  when scaling to 64 chiplets. In contrast, Flat-B delivers the strongest and most consistent scaling doubling the IPC of the Mesh and 2D-Torus and achieving a tenfold improvement over the Ring at high chiplet counts. Most importantly, Flat-B achieves comparable performance to a monolithic design with equivalent compute capability. This demonstrates that disaggregation is key to retaining monolithic performance, given a sufficiently scalable network.

Regarding application-level scalability, three distinct behaviors can be identified. These are largely influenced by the memory access patterns of the workloads which, in turn, define their traffic characteristics. BFS (Figure \ref{fig:sub_ipc_growing_bfs}) is known for its irregular memory access pattern, which generates significant traffic and limits the benefits of additional compute and memory resources, resulting in modest IPC gains beyond 16 chiplets. This result shows that scale-out schemes alone cannot overcome workload-dependent communication bottlenecks, highlighting the need for architecture-aware optimizations to sustain performance as the number of chiplets increases.

FFT (Figure \ref{fig:sub_ipc_growing_fft}) occupies an intermediate position: although it generates substantial communication traffic, its more structured access pattern allows it to benefit significantly from scalable interconnects, delivering twice the performance of the monolithic design with Flat-B at 64 chiplets. Under the same configuration, 2D-Torus and Mesh degrade performance by 46\% and 64\% relative to monolithic, respectively. 

\begin{figure}
    \includegraphics[]{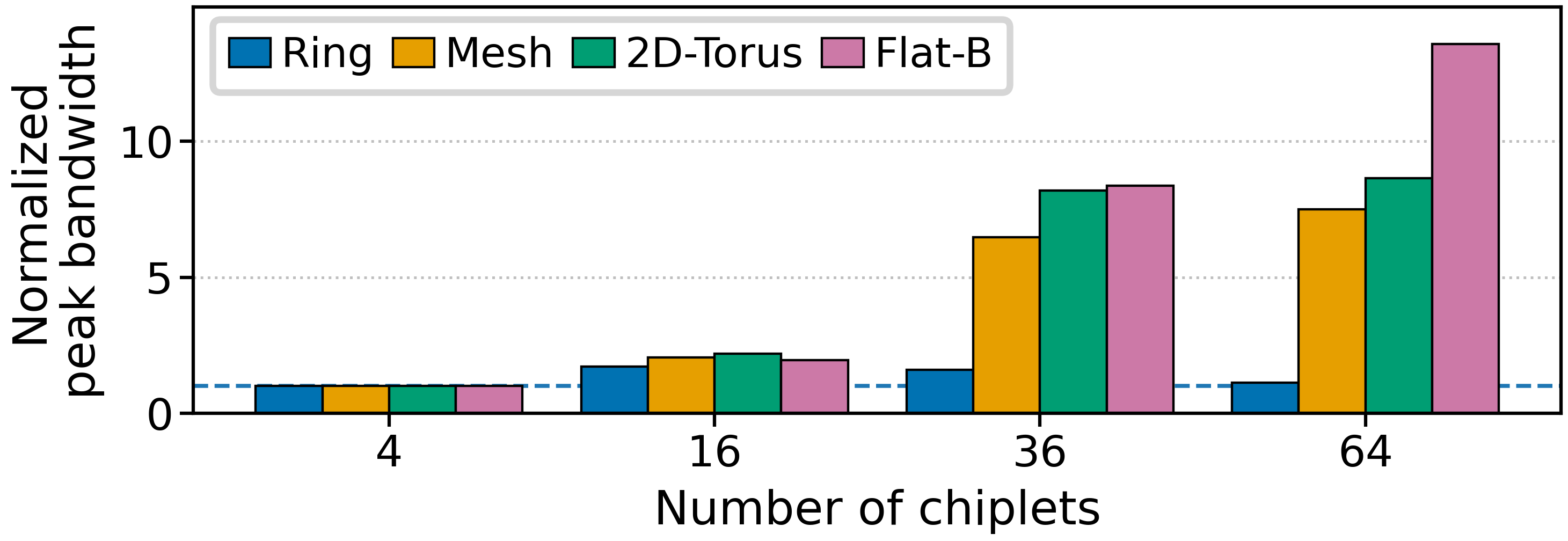}
    \caption{\centering GEMM peak bandwidth.\\ Normalized to 4 chiplet ring topology}
   \label{fig:sgemm_peak_band}
\end{figure}

Finally, GEMM exhibits superlinear scaling under the Flat-B topology, exceeding the linear reference by 66\% and 50\% in the 36 and 64 chiplet configurations, respectively. This result matches monolithic performance closely, which indicates that the superlinear effects can be attributed to memory hierarchy phenomena unrelated to disaggregation, such as better cache locality at higher SM and L2 bank counts, which benefit disaggregated and monolithic designs equally. Figure \ref{fig:sgemm_peak_band} reports the peak bandwidth for GEMM across all topologies and configurations, normalized to the four chiplet ring baseline. Flat-B exhibits the strongest scaling, reaching approximately $13.5\times$ the baseline bandwidth with 64 chiplets, compared with about $7.5\times$ for the Mesh and $8.7\times$ for the 2D-Torus. This higher bandwidth allows Flat-B to sustain the communication demands of GEMM at large chiplet counts, explaining its superior scalability. Together, these results show that the impact of interconnect design grows with communication demand, making scalable topologies increasingly important for communication-intensive workloads.

\textit{\textbf{Finding:}} Scale-out performance in MCM GPUs is constrained by the scalability of the interconnection network, which becomes the determinant of performance as the chiplet count increases. Therefore, only highly scalable topologies can sustain performance at chiplet counts beyond the current four chiplet industry standard, matching or even surpassing a hypothetical equivalent monolithic GPU.

\subsection{Scale-up and Hybrid Scaling}

\begin{figure*}
\centering
\includegraphics[width=\textwidth]{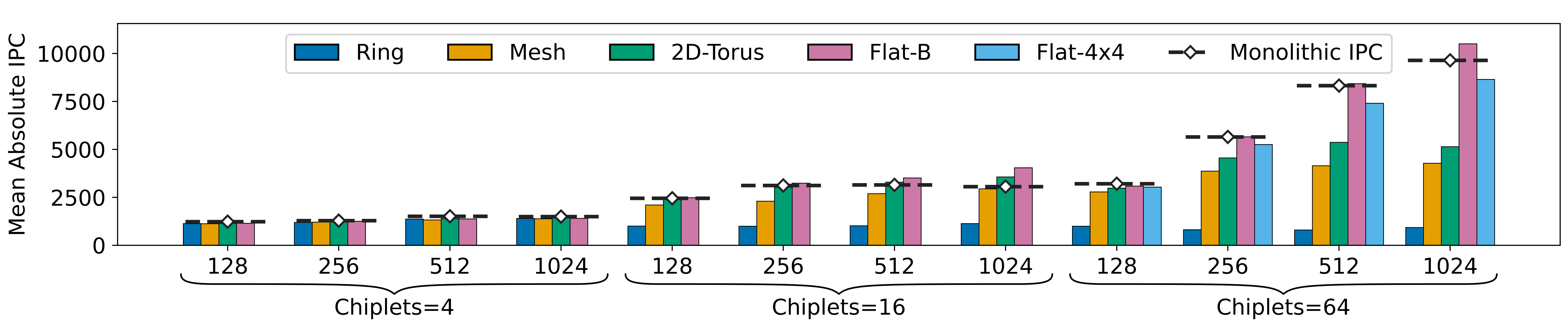}
\caption{Average IPC with varying SM and chiplet counts.}
\label{fig:fixed_gpu_size_avg}
\end{figure*}

Having analyzed the scale-out approach to sustaining MCM GPU performance, we now evaluate the impact of increasing the compute capability of individual chiplets. By considering different chiplet counts, this analysis also captures hybrid scaling strategies that combine scale-up and scale-out mechanisms. Figure \ref{fig:fixed_gpu_size_avg} shows the average IPC achieved by GPUs with different total numbers of SMs, implemented using varying amounts of chiplets and interconnect topologies. Note that, similarly to the scale-out analysis, each chiplet is assumed to feature one memory controller. For each GPU configuration, the black horizontal marker indicates the IPC of a monolithic GPU with equivalent compute and memory resources. This monolithic design is included as a reference point, although it may be impractical or physically infeasible to manufacture for the largest GPU configurations. Additionally, the 64 chiplet design point includes the Flat-4x4 topology, analyzed as an affordable alternative to the fully-fledged Flat-B.

As shown by the results, all the evaluated topologies are close to the performance of a monolithic design when the system is disaggregated into four chiplets. This outcome is not particularly surprising given the short communication distances in networks of this size. However, as the number of chiplets increases, topology becomes a key determinant of scalability, with communication emerging as the main bottleneck. The Ring standard scales poorly, dropping below 50\% of monolithic performance at 16 chiplets systems and below 10\% at 64 chiplets. At 16 chiplets, the Mesh retains about 80\%, while the 2D Torus and Flat-B match or exceed the monolithic baseline. Configurations outperforming the monolithic desing reduce contention by distributing memory requests across smaller chiplets, fewer SMs compete for the same local structures, increasing throughput and potentially offsetting inter-chiplet communication overhead.

However, at 64 chiplets, only Flat-B retains monolithic performance. Its low diameter and high connectivity provide sufficient bandwidth to avoid the severe degradation observed in the Ring, Mesh, and 2D-Torus. The Flat-4x4 results suggest an interesting tradeoff between performance and cost, remaining scalable as the SM count increases and retaining a significant portion of monolithic performance at a fraction of the cost. Moreover, Flat-4x4 outperforms the 2D-Torus, which is more costly in terms of both router and link count.

Overall, except for the Ring, configurations with at least 16 chiplets outperform their four chiplet counterparts showing that greater disaggregation is beneficial when supported by a sufficiently scalable interconnect.

Continuing the analysis of scale-out and scale-up, Figure \ref{fig:fixed_gpu_size_avg} reveals additional opportunities for hybrid scaling. Notably, all 64 chiplet configurations using non-ring topologies, with 256 or 512 SMs outperform the 16 chiplet, 1024 SMs configuration with speedups of up to $1.39\times$ and $2\times$, respectively. Thus, disaggregation can more than compensate for having half or even one quarter of the compute resources, by providing higher aggregate network and memory bandwidth. The implication is profound: performance is not dictated only by the number of SMs, but by the ability of the system to sustain efficient data movement among distributed compute and memory resources. Moreover, the 16 chiplet 256 SMs design delivers $2.58\times$ the performance of a four-chiplet MCM architecture with equivalent compute capability. Therefore, disaggregation and highly scalable topologies will play a key role in the co-design of compute, memory, and inter-chiplet networks, enabling next-generation GPU designs featuring around 256 SMs.

These results also highlight the impact of additional per-chiplet SM pressure on network saturation. In the 64 chiplet configuration, as the total number of SMs increases, the additional network traffic causes increasing congestion in the Mesh and 2D Torus, reducing effective memory access throughput and causing performance to diverge further from the monolithic baseline. For instance, in the largest evaluated design, the Torus achieves only about 50\% of monolithic performance, whereas the Flat-B even surpasses it, given its greater bisection bandwidth and lower hop-count. 

Overall, Figure \ref{fig:fixed_gpu_size_avg} also underlines the limited ability of scale-up approaches alone to sustain scalable performance. A closer examination of the 16 chiplet configuration reveals that monolithic performance is achieved across all evaluated SM counts. Nevertheless, system performance remains largely unchanged as the number of SMs increases, indicating the access to data has become the bottleneck. In contrast, the 64 chiplet configurations exhibit a sustained increase in performance with growing SM counts. This gain could be attributed solely to the additional memory controllers available in the 64 chiplet experiments. To isolate the impact of the inter-chiplet network, additional experiments were conducted in which the 1024 SMs, 16 chiplet configuration was enhanced with four memory controllers per chiplet, effectively matching the compute and memory capabilities of the largest evaluated 64 chiplet configuration. Despite the substantial increase in memory bandwidth, performance improved by only 11\%, remaining far below that achieved by the corresponding 64 chiplet design. These results further emphasize the importance of scale-out to realize the full performance potential of the available resources. Merely increasing resource density is insufficient if the inter-chiplet communication infrastructure becomes a bottleneck. Instead, adequate network scalability is essential to unlock the full performance potential of highly disaggregated GPU architectures.


\textit{\textbf{Finding:}} Network scalability, rather than compute density, is the driver of performance in MCM GPUs. Provisioning additional SMs or memory controllers alone delivers limited benefit, whereas highly scalable topologies sustain monolithic performance even at large chiplet counts. Future MCM GPUs must co-dimension compute, memory, and interconnect, with network design having a major impact on performance.

\subsection{Disaggregation and energy}


\begin{figure*}
\centering
\includegraphics[width=\textwidth]{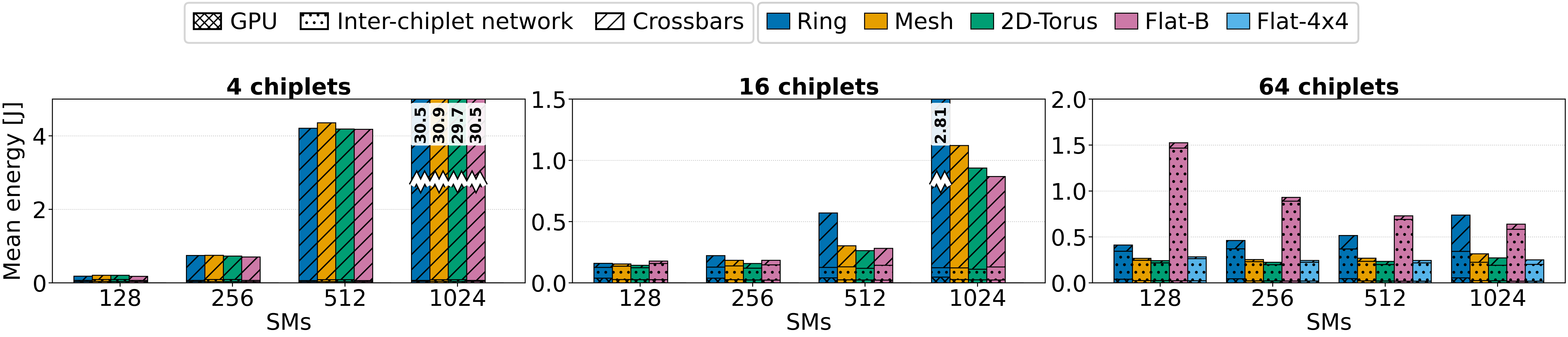}
\caption{Average energy consumption for different SM and chiplet counts.}
\label{fig:fixed_chiplet_counts_energy}
\end{figure*}

\begin{figure*}
\centering
\includegraphics[width=\textwidth]{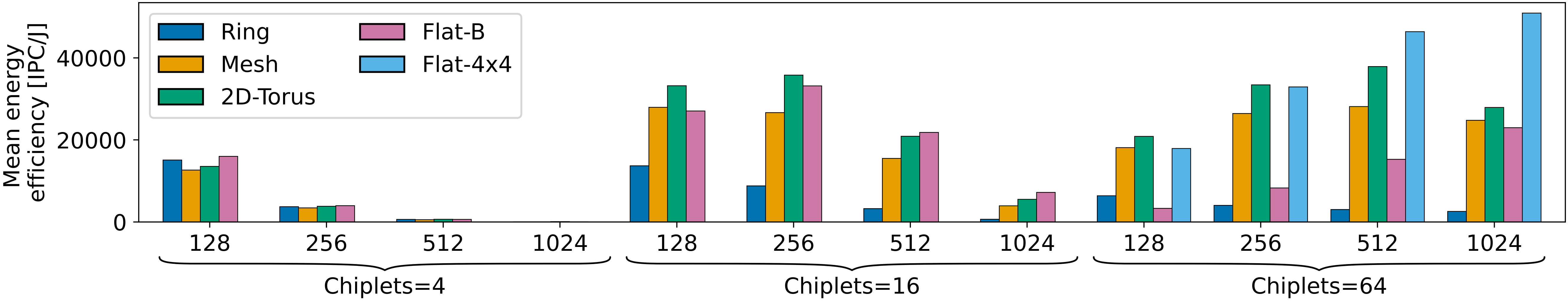}
\caption{Average efficiency for different SM counts under varying degrees of disaggregation.}
\label{fig:fixed_chiplet_counts_eff}
\end{figure*}

Fig.~\ref{fig:fixed_chiplet_counts_energy} reports the total average energy consumption across topologies for different chiplet counts and SM configurations, decomposed into GPU, inter chiplet network, and crossbar contributions. The energy values for the monolithic GPU are omitted because they are excessively large and would compromise the readability of the figure. They are $4.82\times$ to $253.37\times$ higher than the corresponding chiplet-based reference, which would excessively expand the y-axis and obscure the differences among the values shown. Results show that disaggregation is highly beneficial in terms of energy consumption and that, for a fixed chiplet count, increasing the total number of SMs consistently leads to higher energy consumption in the 4 and 16 chiplet configurations. For instance, in the four chiplet configuration, increasing the number of SMs from 128 to 1024 results in an approximately two-order-of-magnitude increase in energy consumption. Notably, this increase occurs despite a 8-fold growth in computational capabilities, which would be expected to substantially reduce execution time and, consequently, mitigate the increase in energy consumption. This result confirms that the crossbar is the dominant energy contributor, particularly once the SM count exceeds approximately 32 SMs per chiplet. A clear example can be observed in the 16 chiplet configuration, in which energy consumption increases sharply when scaling from 512 to 1024 SMs. Current commercial GPU designs operate close to this boundary, highlighting the importance of studying disaggregation and interconnect topologies for future architectures. More importantly, these results suggest that highly disaggregated designs with fewer SMs per chiplet may offer a favorable balance between performance and  efficiency by reducing the cost of the internal crossbar. This is underscored by the significant overall reduction in energy consumption when comparing the 4 and 16 chiplet configurations.

From 16 to 64 chiplets, total energy remains relatively stable for most network topologies. At these levels of disaggregation, each chiplet contains fewer resources, substantially reducing the size and energy contribution of its internal crossbar. Consequently, the inter-chiplet network becomes the dominant source of energy, and the overall efficiency depends increasingly on the balance between network complexity and application performance. Flat-B exhibits a distinctive behavior at 64 chiplets: although its highly connected structure increases network energy, the substantial performance improvement shown in Fig.~\ref{fig:fixed_gpu_size_avg} shortens execution time enough to compensate for much of this additional cost. Additionally, given its exceptional scalability, additional SMs provide further execution time reductions that show in decreasing energy consumption. Flat-$4\times4$, in contrast, provides a more energy efficient design point by reducing the number and complexity of routers, ports, and links. This concentrated organization substantially lowers network energy with respect to the full Flat-B topology, obtaining comparable results to a 2D-Torus.

This is further explored in Figure \ref{fig:fixed_chiplet_counts_eff}, which shows the efficiency in terms of IPC/J (higher is better). Monolithic results are omitted, as their values are between $4.46\times$ and $246.87\times$ lower than those of the corresponding chiplet-based configurations, making them the least energy-efficient option. Contrary to the performance results discussed previously, energy efficiency favors the Torus topology in most evaluated configurations. The Flattened Butterfly only becomes the most efficient option when concentration is considered (Flat-$4\times4$) and for large SM counts. These results reveal potentially attractive design tradeoffs, where different topologies may be preferred depending on whether the primary objective is maximizing performance or improving energy efficiency.

\textit{\textbf{Finding:}} Chiplet-based designs are crucial to the energy viability of large scale GPU designs. A 16 chiplet Torus emerges as the most balanced design point for next-generation GPUs with 256 SMs, striking a balance between cost, performance, and energy. Concentration delivers further efficiency gains for future designs pushing the boundaries of disaggregation.

\section{Related Work}
\label{sec:related_work}

Prior work has explored the design and scalability of inter-chiplet networks in multicore and general-purpose chiplet-based systems. Jerger et al. \cite{jerger2014noc} and Kannan et al. \cite{kannan2015enabling} studied how silicon interposers can support communication between disaggregated CPU chiplets, showing that network organization is critical to preserving performance. Other works proposed scalable interposer topologies and routing mechanisms, including ClusCross \cite{shabani2019cluscross}, Kite \cite{bharadwaj2020kite}, HexaMesh \cite{iff2023hexamesh}, and modular deadlock-free routing \cite{yin2018modular}. More recently, Mallya et al. \cite{mallya2025performance} analyzed the impact of chiplet granularity, network bandwidth, cache organization, and interposer technology on performance. Although these studies mainly target CPUs and general-purpose systems, they reveal communication and scalability challenges that are also relevant to MCM GPUs.

Prior work across MCM GPUs, multi-GPU systems, and CPUs has largely optimized specific components or mechanisms. LADM \cite{khairy2020locality} and CODA \cite{coda} co-locate threads and data through compile-time analysis or address-mapping optimizations. GRIT \cite{wang2024grit} dynamically selects first-touch placement, migration, or replication in multi-GPU systems, while CLAP \cite{park2025leveraging} exploits locality by selecting suitable page sizes and mapping contiguous pages to the chiplet most likely to access them. CARVE \cite{numaMultiGPU} caches remote data in multi-GPU systems, and Memplex \cite{memplex} applies per-chiplet DRAM caching, migration, and replication to multi-chiplet architectures. For CPUs, ARCAS \cite{arcas} places tasks according to cache affinity and inter-chiplet latency. SAC \cite{sacsharing} reconfigures cache placement and routing at runtime, whereas NearFetch \cite{nearFetch} caches data along memory-request paths for opportunistic forwarding. Although these studies optimize locality within specific architectural domains, our work complements them by introducing the idea that dividing compute and memory resources across several chiplets is essential to achieve scalable GPU performance.

\section{Conclusions}
\label{sec:conclusion}


This work demonstrates our initial divide and conquer hypothesis, confirming that disaggregation is an effective architectural strategy for improving the scalability, performance, as well as energy consumption and efficiency of MCM GPUs. However, its benefits depend not only on increasing the number of chiplets, but also on, the scalability strategy and providing an interconnection network capable of sustaining the resulting communication demands.

Our systematic evaluation of scale-up, scale-out, and hybrid strategies shows that network scalability becomes the primary determinant of performance as chiplet count increases. Increasing the number of SMs or memory controllers in isolation provides limited benefits when inter-chiplet bandwidth, and contention constrain resource utilization. In contrast, highly scalable topologies enable disaggregated configurations to surpass equivalent monolithic designs, even beyond the current four-chiplet industry standard.

The results indicate that compute, memory, and interconnect resources must be co-designed and scaled jointly. Among the evaluated configurations, the 16 chiplet Torus with 256 SMs, representative of next-generation GPUs, provides the most balanced design point, delivering $2.40\times$ the performance of a state-of-the-art MCM architecture with the same compute capability while reducing energy consumption by $4.45\times$.
Further disaggregation can yield additional efficiency gains when supported by an adequate network.

Overall, these findings establish disaggregation and interconnect scalability as first-order considerations in MCM GPU design and provide practical architectural guidance for developing scalable, high-performance, and energy-efficient next-generation GPUs. Future work will extend this study by exploring adaptive routing mechanisms, evaluating heterogeneous interconnects that combine different link capabilities, and developing advanced data-placement algorithms to improve locality and reduce inter-chiplet communication.




\bibliographystyle{IEEEtranS}
\bibliography{refs}

\end{document}